\documentstyle[11pt,aaspp4]{article}

\lefthead{Richmond et al.}
\righthead{The Amateur Sky Survey}

\begin {document}

\title{The Amateur Sky Survey Mark III Project}
\author{Michael W. Richmond\altaffilmark{1},
        Thomas F. Droege,
        Glenn Gombert\altaffilmark{2},
        Michael Gutzwiller,
        Arne A. Henden\altaffilmark{3},
        Chris Albertson,
        Nicholas Beser\altaffilmark{4},
        Norman Molhant,
        and 
        Herb Johnson\altaffilmark{5}}
\authoraddr{Physics Department, 85 Lomb Memorial Drive, 
Rochester, NY 14623-5603}
\altaffiltext{1}{Physics Department, Rochester Institute of Technology, Rochester, NY 14623-5603}


\altaffiltext{2}{Miami Valley Astronomical Society}


\altaffiltext{3}{U.S.R.A./US Naval Observatory, Flagstaff Station, Flagstaff, AZ 86002}


\altaffiltext{4}{Applied Physics Lab, Johns Hopkins University}


\altaffiltext{5}{New Jersey Astronomical Association, High Bridge NJ 08829}

\begin {abstract}
The Amateur Sky Survey (TASS) is a loose confederation of amateur
and professional astronomers.
We describe the design and construction of our Mark III 
system, a set of wide-field drift-scan CCD cameras 
which monitor the celestial equator down to thirteenth magnitude
in several passbands.
We explain the methods by which images are gathered,
processed, and reduced into lists of stellar positions and magnitudes.
Over the period October, 1996, to November, 1998,
we compiled 
a large database of photometric measurements.
One of our results is the {\it tenxcat} catalog,
which contains measurements on the standard Johnson-Cousins system
for 367,241 stars;
it contains links to the light curves of these stars as well.
\end {abstract}

\section { Introduction }

In the year 1781, a musician and amateur astronomer named William Herschel 
discovered the planet Uranus while sweeping the skies with a homemade
telescope in the back yard of his home in London.
At that time, wealthy gentlemen could and did possess the equipment
and knowledge to make significant advances in the science.
During the next century, as astronomy evolved into a profession,
academic and research institutions gradually dominated the field:
only they possessed the resources to construct ever-larger telescopes
and equip them with sophisticated instruments.
Observatories moved to remote mountaintops in search of dark skies
and good seeing, leaving the common man far behind.
By the middle of the twentieth century, there were very
few opportunities for amateur astronomers to contribute to the 
discipline: monitoring bright variable stars and atmospheric features of
the planets, certainly, but little else.

In the past two decades, however, fortuitous developments in technology
have given amateur astronomers a chance to rejoin the field.
The combination of inexpensive CCD detectors and cheap, powerful computers
permits any motivated individual to measure quantitatively the 
position and brightness of tens of millions of celestial sources.
While the realm of the nebulae may still belong largely to the
professional, the nearby universe is open to all.

In this paper, we describe a project
in which we constructed our own equipment and
used it to conduct a
survey of bright stars near the celestial equator.
Because most of us are amateurs, we call the project
``The Amateur Sky Survey,'' or ``TASS'' for short.
After several experiments in designing celestial cameras
(codenamed Mark 0 through Mark II), 
we settled on the Mark III device, which
we describe in Section 2.
In Section 3, we characterize the software used to reduce 
our images into lists of stars.
We provide a few details on the observing locations (our back yards)
in Section 4.
We discovered that the really difficult part of this survey
was keeping track of all the data (a common lesson in the current
age of large-scale surveys), and so we devote Section 5 to the
details of our solution.
In Section 6 we discuss several results from the first three
years of our work,
and conclude in Section 7 by describing our future plans.

\section { Hardware }

The Mark III cameras were designed and constructed by one of us (TD)
in 1995,
then distributed gratis to observers across the country.
We arrange our description of the subsystems to follow the order
in which an incoming photon encounters them.

Each camera has an Aubell 135-mm f/2.8 camera lens.
We chose this lens because it was on special at a New York camera
store for \$19,
and was available in sufficient quantity to provide identical
optics to all our units.
We later noticed that while the lens gives reasonably
sharp images (FWHM $\sim 23$ microns) in $V$-band, 
it performs poorly in $I$-band:
the PSF shows a core surrounded by a halo, and its FWHM ($\sim 33$ microns)
is considerably larger than that of the $V$-band PSF.  
We believe this inability to focus near-IR light accounts for
most of the increased scatter in $I$-band photometry.
Our future cameras will feature better optics.

Between the lens and the detector, we place filters manufactured
to the 
Bessell \markcite{bess90} (1990) 
prescription
by Omega Optical, Inc. 
Each Mark III ``triplet'' contains three cameras mounted
together (Figure 1).
We usually chose two $I$-band and one $V$-band filters,
but the Batavia triplet has one each of $V$, $R$ and $I$.

The heart of each camera is a Kodak KAF-0400 CCD detector.
The KAF-0400 has 512 rows and 768 columns of $9$ micron pixels,
with a full-well capacity of 85,000 electrons per pixel.
We operate the CCDs in drift-scan mode: 
each chip is mounted
so that its columns run east-west, parallel to the motion of
stars across the sky.  We read out the chip continuously,
shifting charge along the columns at the same rate that stars
drift across the detector.
For a detailed description of drift-scanning, see
Gehrels \markcite{gehr86} et al (1986).
Our 135-mm lens yields a plate scale of about $13.8$ arcseconds
per pixel;
the effective exposure time is the time required for a star 
to drift 
$512 \times 13.8 {\rm \ arcsec} \sim 1.96$ degrees on the equator:
471 seconds.

The length of our exposures requires us to cool the CCD well below
ambient temperatures.
Each CCD is attached to a two-stage thermoelectric (TE) cooler:
the first stage does most of the work, dropping the temperature
by about $30^{\circ}$ Celsius.
The second stage is driven by a circuit which acts to maintain
the temperature at a constant value.
We place a large bucket of water next to the triplet at 
each site to act as a heat sink at roughly constant
temperature, and circulate the water through the TE coolers.
Although the water may cool down during the course of a night
by a large amount,
the temperature regulation can hold the CCD at a 
temperature constant to within $\sim 0.01^{\circ}$ C.

The specifications of the Kodak KAF-0400 show a read noise of
13 electrons.  When running our cameras at a typical operating
temperature of $-15^{\circ}$ Celsius, we measure a total noise of
25 to 30 electrons in dark images.
Photon noise from the sky at our sites overwhelms the readout
and thermal noise: the total noise ranges
from 50 to 120 electrons per pixel in a $V$-band image,
and from 90 to 150 electrons in an $I$-band image.

Three cameras are bundled together in each triplet (see Figure 1).
The body holds the cameras in a common plane, but points them
at 15-degree intervals.
We set the triplets to point near the celestial equator, with
one camera looking on the meridian (due south),
one looking an hour to the east, the third looking an hour
to the west.
A star may drift through all three cameras during a single night.
We built a single electronics board, located in the back of
the triplet housing, to control all three cameras.
For simplicity's sake, we used a single clock driver to time
the control signals to all three CCDs.
The camera operator at each site tuned the driver 
by trial and error to achieve the best setting.
We discovered that the single driver was a mistake: 
small variations in the focal lengths of the cameras 
made it impossible to optimize the PSF in all three simultaneously.

Each camera produces a long, continuous image of the night sky,
about three degrees wide from north to south and as long
as the night is clear.
The electronics board within the triplet housing sends
the data, one row at a time, over a byte parallel line to a nearby computer.
The computer reads the information and writes it to disk
in 16-bit integer FITS format, breaking the stream of pixels
into images of roughly 900 rows (about 3.5 degrees in Right Ascension).

\section { Observations }

Since the Mark III cameras require power, cooling water, and a
computer, as well as manual intervention to cover them from the
elements, we placed them in our backyards;
the camera operators have day jobs and cannot drive long distances
to dark sites on a daily basis.
Operation consists of uncovering the cameras
at dusk and starting a computer program.  
The operator then leaves the unit alone,
just making sure to cover up the camera before it rains or the
sun shines into the lens.  
In the morning, the data files are processed to
star lists when time is available.

The data presented herein were taken at three suburban sites:
near Cincinnati, Ohio, longitude 84.58 West, latitude 39.20 North,
operated by MG; 
near Dayton, Ohio, longitude 83.87 West, latitude 39.80 North, 
operated by GG; 
and near Batavia, Illinois, longitude 88.33 West, latitude 41.83 North,
operated by TD.
A fourth triplet at the Applied Physics Lab of Johns Hopkins University,
longitude 76.88 West, latitude 39.15 North, has also run regularly.

The weather at these sites is poor, especially in the
winter months:
our coverage of the equator is weakest between 6 and 12 hours
Right Ascension.
Even when the weather is good, we are sometimes unable to collect
data due to other commitments.
As a result, over the roughly two years from October, 1996, 
to November, 1998,
the three sites combined to submit measurements from 175 nights.
Obviously, we could increase the data rate by a factor of two to four
by placing a triplet within a weather-proof housing at a site with
good weather and automating its operation.
We plan to put the first of our next generation of instruments
in Arizona for exactly this reason.

The skies of our suburban locations are 
much brighter than those at most observatories.
On a typical night in Batavia, the sky brightness is about eighteenth
magnitude per square arcsecond in both $V$-band and $I$-band.
At a dark site ($\sim 21.0$ mag/square arcsecond in $V$), 
the noise background due to sky brightness would decrease by
a factor of about four, which would increase our limiting magnitude
by a bit more than one magnitude.

During an eight-hour night, each camera scans roughly 360 square
degrees.  The images are stored on disk and processed into 
star lists the next day.
Depending on the season and sky brightness, a triplet
may record 50,000 to 200,000 measurements 
on a good night.
Most sources are detected two or three times, at one-hour intervals.
Our images provide good data for 
sources from seventh magnitude to thirteenth or fourteeth magnitude.
We provide an example of a $V$-band Mark III image in Figure 2:
it shows a field roughly 4 degrees wide by 3 degrees high
at the intersection of Monoceros, Canis Minor, and Hydra.
Figure 3 shows a closeup view of the area bounded by the
dashed box in Figure 2;
note the slightly diagonal shape of the PSF.

\section { Software }

We have written all the software used to record the image data,
clean the images, detect and measure stellar sources.
In each step, we follow standard procedures and employ conventional
algorithms; nonetheless, we describe our pipeline in some detail
so that readers may understand the limitations it places on
the final results.

\subsection { Cleaning the raw images }

Since we take drift-scan images, the signal from each star passes
through every row along one column of the CCD.  
Correcting the images requires the construction and application
of one-dimensional
dark and flatfield vectors.

We make a dark vector for each camera on each night by taking an
image of several hundred rows
with the lens cap in place, then calculating the median of the 
pixel values in each column.

We create a flatfield vector in a slightly different manner.
In order to improve the signal-to-noise ratio, we start with 
scans taken during twilight.  
First, we subtract the dark vector from the scans.
Now, since the brightness level changes 
rapidly at these times, we cannot simply take the median of several
hundred rows as a fair measure of the change in sensitivity 
across the chip.
Therefore we calculate the flatfield vector in two steps.  First, we create
a set of temporary flatfield vectors, one for each row in our twilight
image, by dividing the pixel values in that row by the median pixel value of
that row.  
We then create a single flatfield
vector whose column values are the median value of each column from our set of
temporary flat vectors.

We normalize the flatfield vector and then divide each
dark-subtracted raw image by it.

\subsection { Generating a PSF }

For each frame, we first calculate the sky value and the standard 
deviation from that value.
Next, we find all sources with peaks at least 20 times the standard
deviation above the sky.
To make a quick estimate of the Full Width at Half-Maximum (FWHM)
of the PSF, we walk down from the peak pixel along the rows and columns 
until the pixel value drops to half that of the peak.
We use the median value of this rough FWHM in the row and column directions
as a first approximation.

We make a second pass through the sources, discarding any with 
FWHM values which deviate from the first approximations by more 
than $25\%$. 
We calculate the average FWHM in each direction from the remaining peaks.
We then create a bounding rectangle whose  width and height are 2.55 times
the average FWHM (i.e. 3 sigma for a true gaussian) in each direction.  
We create a set of PSF's by normalizing each peak within the bounding
rectangle.  
We average all the PSF's in the set to determine the final PSF.
We therefore have a strictly empirical PSF for each image.

We build an ``optimized aperture'' from this discrete PSF by
selecting whole pixels, working outwards from the center, until
the next pixel value drops to less than one-fourth of the
average pixel value included so far.
This aperture has the same shape as the PSF, but discards
pixels in which the signal-to-noise of a star is low.

\subsection { Finding stars }

Following DAOPHOT 
(Stetson \markcite{stet87} 1987),
we convolve the image with a lowered PSF to detect sources.
We mark as a candidate peaks in the convolved image 
above a threshold (typically two or three times the standard
deviation from the sky).
We measure the sky-subtracted flux of the candidate in 
the original image
through the optimized aperture, 
and use it to calculate the ratio 

\begin{displaymath}
f = { \rm { flux\ inside\ aperture } \over  
         { \rm { peak\ in\ convolved\ image } } }
\end{displaymath}

A true star has a value of $f$ close to 1.0,
while cosmic rays and other spurious detections typically
yield very different values.
We accept any candidate with $f$ within a reasonable range
of the expected value.

\subsection { Measuring stellar properties }

To calculate the position of each detected star, we 
use pixels within the aperture to form marginal sums
in the row (column) directions.
We add the sums for each row (column) to form a cumulative
marginal sum, then interpolate to find
the row (column) at which the sum reaches half its final value.
Our interpolation scheme yields positions which
are fractions of a pixel.

We convert the (row, col) position of each star into (RA, Dec)
by matching the brightest 30 stars in each image against
stars in the Tycho catalog 
(ESA \markcite{hipp97} 1997);
we have adapted the triangle-based method described 
by 
Groth \markcite{grot86} (1986)
and
Valdes et al. \markcite{vald95} (1995)
to match the detected and catalog positions of the stars.
Each detection is tagged with the time at which it passed
the midpoint of the field of view, based on its
position within the image and the time at which the image
was read from the camera.

To calculate the flux of each detected star,
we do not use this interpolated position.
Instead, we adopt the position of the peak in the
image convolved with the PSF.
At this position in the original image
we place the ``optimized aperture,''
and add (flux - sky) contributed by each pixel within the aperture.
We estimate an uncertainty in each measurement by combining 
contributions from readout noise, sky subtraction, and photon
noise from the source itself.

\subsection { Photometric calibration }

Each Mark III camera converts instrumental magnitudes to the
standard Johnson-Cousins system via a two-step calibration procedure.

First,
in order to set a preliminary zero-point for the instrumental
magnitudes, we consider a particular subset of stars from the
Tycho catalog.
We choose stars which 
\begin{itemize}
\item are not marked as variable in the Tycho catalog,
\item have $B_T$ and $V_T$ uncertainties of $\leq 0.06$ mag,
\item are fainter than $V_T = 7.5$ so that they are not saturated,
\item are separated from other Tycho stars by at least 83 arcsec in RA
          and 50 arcsec in Dec.
\end{itemize}
Since the TASS cameras observe in Johnson-Cousins $VRI$, we need
to convert the Tycho $B_T$ and $V_T$ magnitudes to the Johnson-Cousins
system.  
We found a set of 300 Tycho stars brighter than $V=9$ which matched
Landolt (\markcite{land83} 1983, 
\markcite{land92} 1992) 
standards.
We performed linear fits to $V, R,$ and $I$ versus $(B_T - V_T)$,
finding transformation equations:
\begin{eqnarray}
      V & = & \phantom{-}0.0156  + V_T  - 0.0994 (B_T - V_T) \\
      R & = & -0.0160  + V_T  - 0.5390 (B_T - V_T) \\
      I & = & -0.0468  + V_T  - 0.9480 (B_T - V_T) 
\end{eqnarray}
Note that the V equation is similar to that given in the
Hipparcos (ESA \markcite{hipp97} 1997) catalog:
\begin{eqnarray}
     V & = & V_T - 0.090(B_T - V_T) 
\end{eqnarray}
but with slightly different coefficients.  To ensure uniformity,
we adopted the new coefficients.  
Note also that each of these
equations is only valid over a limited color range, 
and the quality of the fit degrades as one moves further from the $B_T$
passband.  We have therefore further restricted the Tycho
subset to stars within the color
range of $-0.2 < (B_T - V_T) < 2.0$.

We then compare these approximate Johnson-Cousins magnitudes 
to our instrumental magnitudes for the same stars in our images;
we simply shift the instrumental magnitudes by a constant
to match the catalog values.
This procedure places our
observations on the standard Johnson-Cousins system without
having to make separate observations of Landolt fields
(often impossible for cameras scanning a few degrees from 
the celestial equator),
and makes use of non-photometric nights since the Tycho reference
stars are contained within each image.

After a year or so of operation, we discovered that cameras at
several sites exhibited small, systematic errors in photometry as
a function of Declination.
Apparently, the one-dimensional flatfield vectors we create 
do not perfectly remove variations in response across the
field of view.
Fortunately, the errors are nearly constant over the course of a single
night.
We correct them by breaking the three-degree Declination range of 
each camera into a small number (typically 8) of zones.
In each zone, we make a linear fit to the (observed - expected)
residuals as a function of Declination, using Tycho stars from the
entire night;
we force the corrections to agree at each boundary between zones.
We then add these corrections to the magnitude of all stars observed
on that night.
This Declination-dependent correction reduces the residuals of
repeated measurements of bright stars in the $V$-band from 
about 55 millimagnitudes to about 35 millimagnitudes.

The results of this two-step procedure are stored locally
by each site and made available for further analysis.
We have detected small color terms in these measurements 
from some of the Mark III cameras at some of the sites;
see the discussion on color corrections in the {\it tenxcat} catalog
in Section 6.2.

\section { Database }

There have been three to four Mark III triplets active at any
time over the past three years.  
During a typical night of operation, a single triplet measures
over 100,000 stellar positions and magnitudes.
Keeping track of the data generated at all sites is not
a trivial task.

The operator of each triplet is responsible for processing the
data into ``star lists:'' ASCII tables of stellar positions and
magnitudes, with a small amount of ancillary information.
Each operator periodically sends accumulated star lists via
FTP, ZIP disks or CDs,
to one of our members (MR) who acts as DataBase Manager (DBM).
The DBM occasionally loads the star lists into our central
database.
In no sense is this a real-time operation: the interval between
observation and insertion into the database varies from weeks to 
months.

We have adopted the PostgreSQL{\footnote{http://www.postgresql.org}} 
database engine for our project:
it is a relational database with an SQL syntax which runs on
many platforms and is distributed free of charge.
One of us (CA) designed the tables required to hold the
information generated by each site, and the software necessary
to load the information into the database.
The central database runs on the desktop computer in the DBM's office,
where it competes for disk space and memory with many other
projects.
The information from roughly 175 nights of observation 
requires about 5 Gigabytes of disk space.

Once the data has been loaded into the database,
it may be analyzed in many ways.
Users from any computer with an SQL client and an Internet connection
can access the information freely.
We have built WWW interfaces to service common queries
so that novice users need only a web browser to answer simple questions.

One of the main functions of the database is to identify
measurements of the same star in different images,
so that one can calculate mean values or look for variability.
We wrote software to perform this merging of data
with the following method in mind:
if a new detection appears more than $D$ arcseconds from any existing
entry in the database, consider it a new entity;
but if it appears within $D$ arcseconds of an existing entry,
mark it as an additional observation of the existing entry.
We chose $D = 15$ arcseconds, based on the precision of our 
measurements of position (see below) and on the large size of our pixels.

We seeded the database with stars from the USNO A1.0 catalog
(Monet \markcite{mone96} 1996)
down to sixteenth magnitude, to serve as initial, accurate positions.
Our plan was to keep the USNO A1.0 position as the database
position until a star was recorded some large number $N$ times;
then we would switch to using the star's mean observed position.
In this way, random errors in the first few recorded measurements
would not cause the position to wander by large amounts 
(several arcseconds), but, eventually, the database would contain
a position based on our measurements alone.
Unfortunately, we made an error in our software, setting $N = 1$;
thus, the position for each star in our database {\bf did}
wander by significant amounts.
In some cases, the database position moved so far from the true 
position that a new measurement did not fall within $D = 15$ 
arcseconds of the existing position;
the new measurement was then recorded (incorrectly) in the database
as an independent star.
We estimate that perhaps 5 percent of all stars in our database
suffer from this ``spurious companion'' problem; 
we provide one post facto fix in 
our discussion of the {\it tenxcat} catalog, below.

\section { Results }

We are still exploring the wealth of information garnered during
two years of operation, and several sites are still contributing
new observations.
The items mentioned below are simply the first projects to which
we have put our energies; 
many avenues of study await the curious researcher.
For example, we have not yet tried to perform any search
for objects which move in the sky from one night to the next.

In one sense, the primary result of our work 
cannot be presented in this paper:
it is the database of measurements we have collected.
This database continues to grow as we incorporate new
observations from several sites.
There is no restriction on access to the data;
TASS members and non-members alike may sift through it on
an equal basis.
We encourage readers who have a use for the information 
to visit the WWW 
pages{\footnote{http://www.tass-survey.org/tass/www\_scripts/make\_chart.html}}
which describe database access.

\subsection { Precision of Astrometry and Photometry }

The Mark III cameras do not provide precise positions.
The telephoto lens and CCD yield a plate scale of about $13.8$
arcseconds per pixel, the PSF is often asymmetric, 
and the $I$-band PSF features an extended halo.
Comparing the position of a single TASS measurement for a bright star 
to the position listed in the Hubble Guide Star Catalog
(Lasker et al \markcite{lask90} 1990),
we find a scatter of about 3 arcseconds;
the scatter rises to about 6 arcseconds near our detection limit.
The mean of many measurements is considerably more
accurate; see Section 6.2 below.

The photometric measurements produced at each camera site
and transmitted to the central database may suffer from
systematic color terms (see discussion in {\it tenxcat} section below).
We ignore that source of error here, and consider only the 
precision of the measurement: as a metric, we use the
standard deviation from the mean of measurements of the 
same star by the same camera on many different nights.
We find that the standard deviation from the mean 
has a minimum of about
0.03 mag for bright stars, 
increasing to 
more than  0.20 mag for stars near the limits of detection.

If one calculates the expected uncertainty in photometric measurements,
using measured values for readout noise and estimates of the sky 
brightness, one finds much smaller values.  
The Mark III measurements are evidently not limited by photon
noise, except for the very faintest stars.
The major sources of error are imperfect sky subtraction
and variations in the PSF across our field of view.

Since we measure most stars many times,
the mean values of our measurements should provide
information on each star more precise than
that from a single observation.
We therefore have constructed a catalog which
contains the mean values of position and magnitude 
for stars observed multiple times.

\subsection { The { \it tenxcat } Catalog }

After two years of operation, the total number of
detections reported to the central database
reached more than 10 million in $V$, 
0.7 million in $R$, 
and 13 million in $I$.
However, many of these detections were due to noise peaks,
airplane trails, cosmic rays, and other contaminants.
In order to generate a catalog of reliable celestial
sources, 
we selected objects in the database which appeared 
on at least 10 different occasions, in any combination
of passbands.
We call this the {\it tenxcat} catalog: 
it contains 367,241 stars at the time of writing (September, 1999).
Interested readers can access the catalog via the Internet at

\centerline{\it http://www.tass-survey.org/tass/www\_scripts/make\_chart.html }

We subjected the objects in this subset to two additional
steps of processing.
\begin{itemize}
\item We checked each of the Mark III 
      cameras for color terms, by comparing their measurements of
      equatorial standard stars against those of Landolt
      (\markcite{land83} 1983; 
       \markcite{land92} 1992).
      The $R$-band and $I$-band cameras had statistically significant
      residuals as a function of stellar color, which we list
      in the Appendix.
      We therefore applied linear corrections to the TASS magnitudes
      from those cameras to bring their measurements to the
      Johnson-Cousins magnitude scale.

\item We checked each star for ``spurious companions:''
      neighbors within $D = 15$ arcseconds which never appear
      in the same image as the star, and have the same brightness 
      to within 0.5 mag.
      Whenever possible, we merged such pairs of stars into 
      a single entry in our database, and
      marked the remaining entry with a flag.
\end{itemize}
      
We are constrained by our drift-scan cameras to observe near the
celestial equator: our surveyed area described a rough band from Declination 
$= -5^{\circ}$ to $1.5^{\circ}$.
We emphasize that this is by no means a complete catalog;
as Figure 4 shows, the number of observations
varies considerably throughout our survey area.
Since some faint stars were not detected in all images,
an area would need to covered many more than 10 times
for all its stars to be included in {\it tenxcat}.

The {\it tenxcat} catalog contains stars ranging from
about seventh magnitude to about fourteenth
magnitude.
However, the distribution of stellar magnitudes (Figure 5)
shows that the number of stars per magnitude bin begins
to fall at about 
$V = 13.3$, $R = 13.3$,
$I = 12.5$. 
The difference between the passbands is a combination
of the color of a typical star and the system sensitivity
as a function of wavelength.

The internal consistency of our photometric measurements 
can be gauged by calculating the median value of the 
standard deviation from the mean within magnitude bins.
In Figure 6 we show the results for all 3 passbands;
all three have a floor of about $\sigma = 0.04$ mag, 
but the scatter increases much more quickly in $I$ band
than in $V$ or $R$.
We blame the $I$-band results on three factors:
first, the $I$-band images suffered from a core-halo PSF;
second, we combined $I$-band measurements from 5 different cameras at
3 different sites, whereas there were only 3 different cameras
with $V$ filters and a single camera with $R$ band filter;
third, our photometric calibration in $I$ depends upon
extrapolation in the transformation from Tycho $B_T, V_T$
to Johnson-Cousins $I$ (see the Appendix).

The positions in {\it tenxcat} are the mean values from 
at least 10 different measurements.
We have compared our positions against those in 
the ACT Reference Catalog 
(Urban et al \markcite{urba98} 1998)
to judge the accuracy of the mean positions.
We find the median difference between TASS and ACT 
positions to rise slowly with magnitude,
from $0.59$ arcseconds at $V = 7.5$ to 
$0.89$ arcseconds at $V = 11.5$.
The mean positions of fainter stars are less accurate,
but still good to a few arcseconds.

In order to
facilitate the identification of sources,
we include cross-references between stars in {\it tenxcat}
and other catalogs.
Each entry in our catalog has a field for the matching
entry in the Hubble Guide Star Catalog 
(Lasker et al \markcite{lask90} 1990)
and the ACT Reference Catalog
(Urban et al \markcite{urba98} 1998).
We have extracted from the ACT Reference Catalog 
information from the Henry Draper (HD) catalog 
(Cannon \& Pickering \markcite {cann18} 1918-1924;
Cannon \markcite {cann25} 1925),
including the spectral type;
unfortunately, {\it tenxcat} contains only 7792 stars with data from
the HD.
Finally, we have compared the position of each of our stars
against sources in the IRAS Point Source Catalog 
(Beichman et al \markcite{beic88} 1988);
any star which falls within 20 arcseconds of an IRAS source
is marked with the IRAS identifier.

The catalog is available as an ASCII text file,
and via a WWW form which allows the user to search through it
in several ways.
A detailed description of each field in the catalog, 
and instructions on accessing it electronically, are
provided in 
Richmond (\markcite {rich99b} 1999b).

\subsection { Variable Stars }

One of the strengths of the Mark III survey is the number of times
it has scanned some areas of the sky.
Recall that two of our sites have two $I$-band cameras on their triplets,
providing measurements of each star roughly two hours apart during
a single night.
In addition, we have spent many nights observing the celestial
equator during the past few years.
As a result, of the 367,241 stars in {\it tenxcat}, 
over 16,000 have at least 30 measurements in $V$-band and
over 55,000 have at least 30 measurements in $I$-band.
We have arranged our database so that remote users may easily
retrieve all observations of one particular star, or generate
a light curve on the fly.

One obvious use for this wealth of data is to study variable stars.
We describe here several different ways in which our members
have started to mine the database.

Gombert 
(\markcite {gomb98a} 1998a; 
 \markcite {gomb98c} 1998c; 
 \markcite {gomb99} 1999)
has searched for variable stars which do not appear in existing
catalogs, finding over 60 candidates.
He finds that the Mark III data is especially well-suited to finding
Mira-like variables: one can isolate stars with large $(V-I)$ colors
and examine several years of observations.
Gutzwiller (\markcite {gutz99} 1999)
has found another 52 candidates.
These efforts are by no means an exhaustive search of the entire
Mark III dataset, but will keep us busy gathering followup observations
(for an example, see 
Dvorak \markcite {dvor99} 1999).

While our positions are not as accurate as those in astrometric catalogs,
they are certainly better than many listed in the General Catalog
of Variable Stars (GCVS)
(Kholopov et al \markcite {khol92} 1992).
When the identity of a variable in the GCVS is uncertain,
due to a poor position,
one can examine the light curve of all stars in the vicinity 
in our database and often find the variable easily.
Gombert (\markcite {gomb98b} 1998b)
provides positions accurate to a few arcseconds for
51 known variable stars.

Our archives do not include images of the sky, only measurements
of position and brightness for the sources we detect.
The Stardial project 
(McCullough \& Thakkar \markcite {mccu1997} 1997),
on the other hand,
does save all of its images,
but doesn't perform any source detection or measurement.
Since the Mark III and Stardial surveys share a very
large area on the sky 
(about 1400 square degrees in a strip from 
Dec = $0^{\circ}$ to $-4^{\circ}$),
one can combine them to find objects 
which would not stand out in either survey alone.
Hassforther and Bastian (\markcite {hass99} 1999)
have found a Mira variable by searching visually 
a set of Stardial images;
although the star disappears from Stardial at minimum light,
it is still detected in our $I$-band measurements.
We expect to see many other projects combine the
Mark III data with information from other sources.

\section { Future Work }

The Mark III survey will continue for several years.
We have already gathered several seasons of data from a fourth
site (the Applied Physics Lab at Johns Hopkins) to be incorporated
into the database.
This paper represents only a ``snapshot'' of a working project;
we may export several updated versions of our catalog(s) before
reaching the final version.

TASS members have shifted most of their attention and efforts
away from the Mark III survey to a new project:
the Mark IV systems.
A Mark IV unit will consist of two cameras fixed side-by-side
on a common mount;
each camera combines a 100mm f/4 lens, a single filter,
and a Lockheed/Fairchild 442A CCD with $2048 \times 2048$ pixels
to provide a field of view four degrees on a side.
The mount allows motion in Hour Angle and Declination,
and is designed to track accurately for several minutes.
We estimate the detection limit of the system to be fainter
than fifteenth magnitude.
We are currently constructing seven Mark IV systems, the first of which
will be deployed at Flagstaff in Fall 1999.

\acknowledgements

TASS members have supported this project with their own time,
energy and money; no tax dollars were spent on our work.
We have been fortunate enough to receive valuable advice
and contributions
from many people, 
including (but not limited to)
Paul Bartholdi,
Bernie Kluga,
Alain Maury,
Peter McCullough,
Peter Mount,
Marty Pittinger,
Jure Skvarc, 
Saral Wagner,
and
Ron Wickersham.
We also owe a large debt to Bohdan Paczynski, 
whose enthusiasm and generosity guided us through
the crucial early stages of the project.
Jim Kern kindly suggested ways to improve this manuscript.

\appendix
\section{Appendix}

There were 23 cameras built and distributed for the Mark III survey.  
Of these, 16 are known to have operated and to have taken at 
least some data.
Twelve have been operated almost every clear night at their locations.
Data from 9 cameras is presently included in database:
\begin{itemize}
\item 3 V-band cameras, one at each site 
\item 1 R-band camera, at Batavia 
\item 5 I-band cameras: 2 each at Dayton and Cincinnati, 1 at Batavia 
\end{itemize}

We compared TASS measurements of stars against 
Landolt's equatorial catalogs of UBVRI
standard stars 
(Landolt \markcite{land83} 1983; 
\markcite{land92} 1992). 
The V-band measurements showed no significant offset in the mean, nor any
significant color dependence. 
We therefore made no correction to the V-band measurements. 
In the R and I bands, however, there were both 
small offsets in the mean (always in the sense that TASS
measurements were slightly fainter than Landolt measurements), 
and small trends in residuals as a
function of (V-I) color. 
See TASS Technical Note 54 
(Richmond \markcite {rich99a} 1999a)
for details, graphs, and tables. 

We determined the following color terms, 
based on stars brighter than mag 11 and using a {\it color}
which was the mean TASS V value (based on measurements from all cameras) 
minus the mean
TASS I value (based on measurements from all cameras). 
In the equations below, uppercase values
represent corrected single magnitude measurements, 
and lowercase values represent raw single magnitude measurements. 

\begin{eqnarray}
                         R-band  \nonumber \\
  {\rm camera\ H1\ (Batavia):\ }  R &= r + 0.01209 - 0.073704*(color) \nonumber \\
\nonumber \\
                         I-band \nonumber \\
  {\rm camera\ B0\ (Dayton):\ }   I &= i - 0.0563  - 0.018676*(color) \nonumber \\
  {\rm camera\ B2\ (Dayton):\ }   I &= i - 0.0683  + 0.040971*(color) \nonumber \\
  {\rm camera\ D0\ (Cinn): \ }   I &= i - 0.0844  + 0.06116 *(color) \nonumber \\
  {\rm camera\ D2\ (Cinn): \ }   I &= i - 0.1001  + 0.08510 *(color) \nonumber \\
  {\rm camera\ H2\ (Batavia):\ }  I &= i + 0.0121  - 0.073704*(color) \nonumber \\
\end{eqnarray}

After making the corrections to each individual measurement, 
we then re-calculated the mean
magnitude of each star in all passbands. 
In cases of spurious pairs, we made the corrections to each
member of the pair separately, 
based on its own color, re-calculated mean magnitudes for each
member separately, and finally determined a weighted 
mean magnitude for the merged entity. 

The corrected {\it tenxcat} magnitudes agree well with Landolt magnitudes, 
with little dependence on color.

\begin {references}
\reference{beic88} Beichman, C. A., Neugebauer, G., Habing, H. J.,
                      Clegg, P. E., \& Chester, T. J. 1988,
                      IRAS Catalogs and Atlases, Version 2.  Explanatory
                      Supplement, NASA Ref. Publ. 1190
\reference{bess90} Bessell, M. S. 1990, \pasp, 102, 1181
\reference{cann18} Cannon, A. J. \& Pickering, E. C. 1918-1925,
                      The Henry Draper Catalogue,
                      Ann. Astron. Obs. Harvard College, 91-99
\reference{cann25} Cannon, A. J. 1925, The Henry Draper Extension,
                      Ann. Astron. Obs. Harvard College, 100, 17
\reference{dvor99} Dvorak, S. 1999, http://home.att.net/~RollingHillsObs
\reference{gehr86} Gehrels, T., Marsden, B. G., McMillan, R. S. \&
                      Scotti, J. V. 1986, \aj, 91, 1242
\reference{gomb98a} Gombert, G. 1998a, Inf. Bull. Var. Stars, 4575
\reference{gomb98b} Gombert, G. 1998b, Inf. Bull. Var. Stars, 4609
\reference{gomb98c} Gombert, G. 1998c, Inf. Bull. Var. Stars, 4653
\reference{gomb99} Gombert, G. 1999, Inf. Bull. Var. Stars, 4709
\reference{grot86} Groth, E. J. 1986, \aj, 91, 1244
\reference{hass99} Hassforther, B. \& Bastian, U. 1999, 
                      Inf. Bull. Var. Stars, 4742
\reference{hipp97} ESA 1997, Hipparcos and Tycho Catalogues, 
                      ESA SP-1200, 1, 57
\reference{khol92} Kholopov, P. N., Samus, N. N., Durlevich, O. V., 
                      Kazarovets, E. V., Kireeva, N. N., \& Tsvetkova, T. M.
                      1992, Bull. Inf. Centre Donnees Stellaires, 40, 15
\reference{lask90} Lasker, B. M., Sturch, C. R., McLean, B. J., 
                   Russell, J. L., Jenkner, H., \& Shara, M. M. 
                   1990, \aj, 99, 2019
\reference{land83} Landolt, A. U. 1983, \aj, 88, 439
\reference{land92} Landolt, A. U. 1992, \aj, 104, 340
\reference{mccu97} McCullough, P., \& Thakkar, U. 1997, \pasp, 109, 1264
\reference{mone96} Monet, D. 1996, USNO A-1.0: a catalog of astrometric
                   standards (Washington, DC: USNO)
\reference{rich99a} Richmond, M. W. 1999a, TASS Technical Note 54, \break
                       http://www.tass-survey.org/tass/technotes/tn0054.html
\reference{rich99b} Richmond, M. W. 1999b, TASS Technical Note 56, \break
                       http://www.tass-survey.org/tass/technotes/tn0056.html
\reference{stet87} Stetson, P. B. 1987, \pasp, 99, 191
\reference{urba98} Urban, S. E., Corbin,  T. E., Wycoff, G. L., 
                   Martin, J. C., Jackson, E. S., Zacharias, M. I.
                   \& Hall, D. M. 1998, \aj, 115, 1212
\reference{vald95} Valdes, F. G., Campusano, L. E.,
                   Velasquez, J. D., \& Stetson, P. B. 1995, \aj, 107, 1119

\end {references}

%
%
\newpage

\begin{figure}
\figurenum{1}
\caption{
A Mark III triplet, seen from the back (North) side.
The lens has been removed from the East camera.
The plastic tubing circulates cooling water through
each camera.
}
\end{figure}

\begin{figure}
\figurenum{2}
\caption{
Mark III $V$-band image of the area near (J2000) 
RA = 08:09:45, Dec = -01:06:30,
covering 4 degrees E-W by 3 degrees N-S.
Star A = HD 66950, B = HD 68667, C = HD 67111, D = HD 67720, E = HD 67719.
North is up and East to the left.
}
\end{figure}

\begin{figure}
\figurenum{3}
\epsscale{0.80}
\plotone{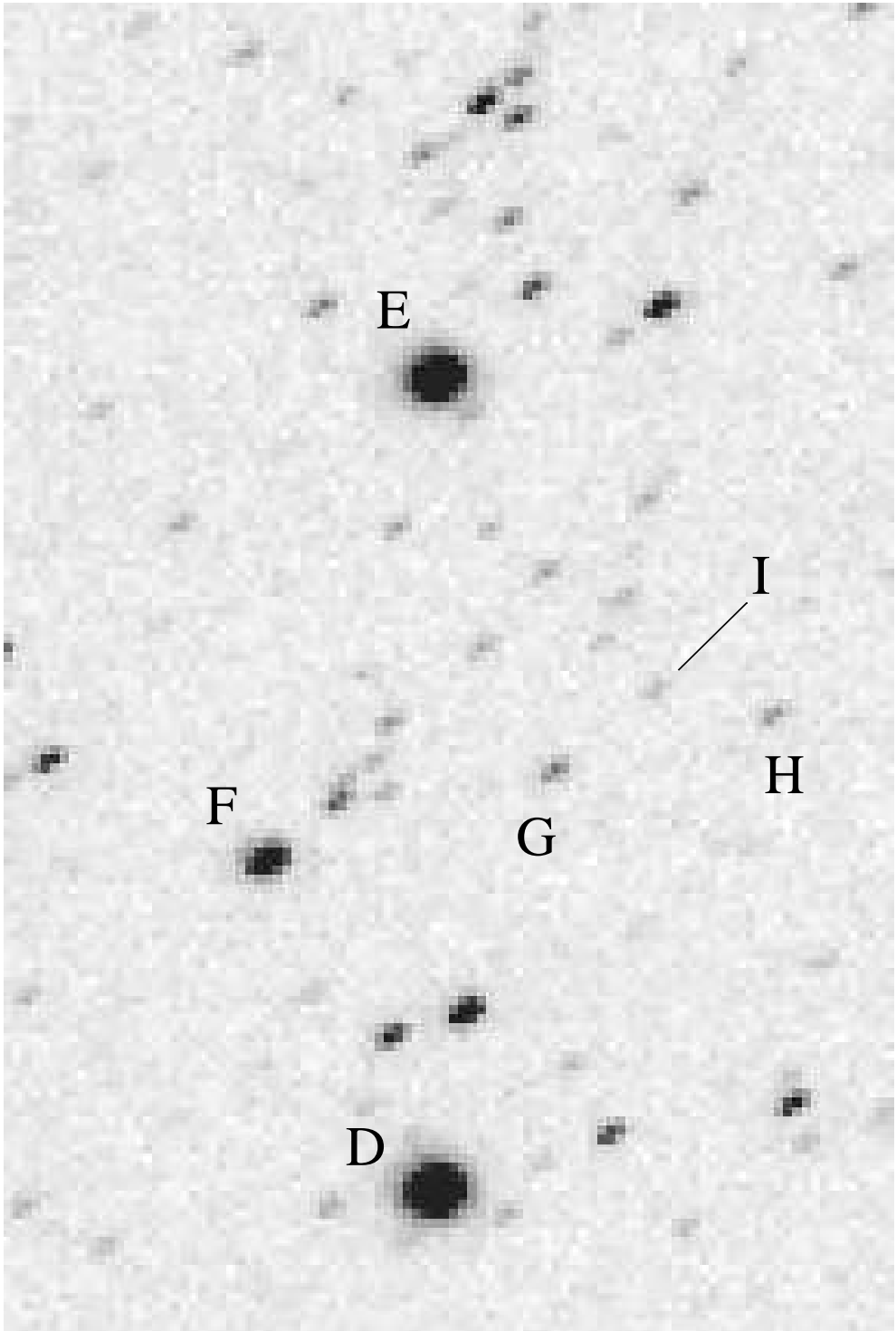}
\caption{
Closeup of the dashed box in Figure 2.
Star F = HD 67769 ($V = 9.4$), G = GSC 4847-02290 ($V = 12.0$), 
H = GSC 4847-02482 ($V = 12.4$), I = GSC 4847-02614 ($V = 13.2$).
North is up and East to the left.
}
\end{figure}

\begin{figure}
\figurenum{4}
\plotone{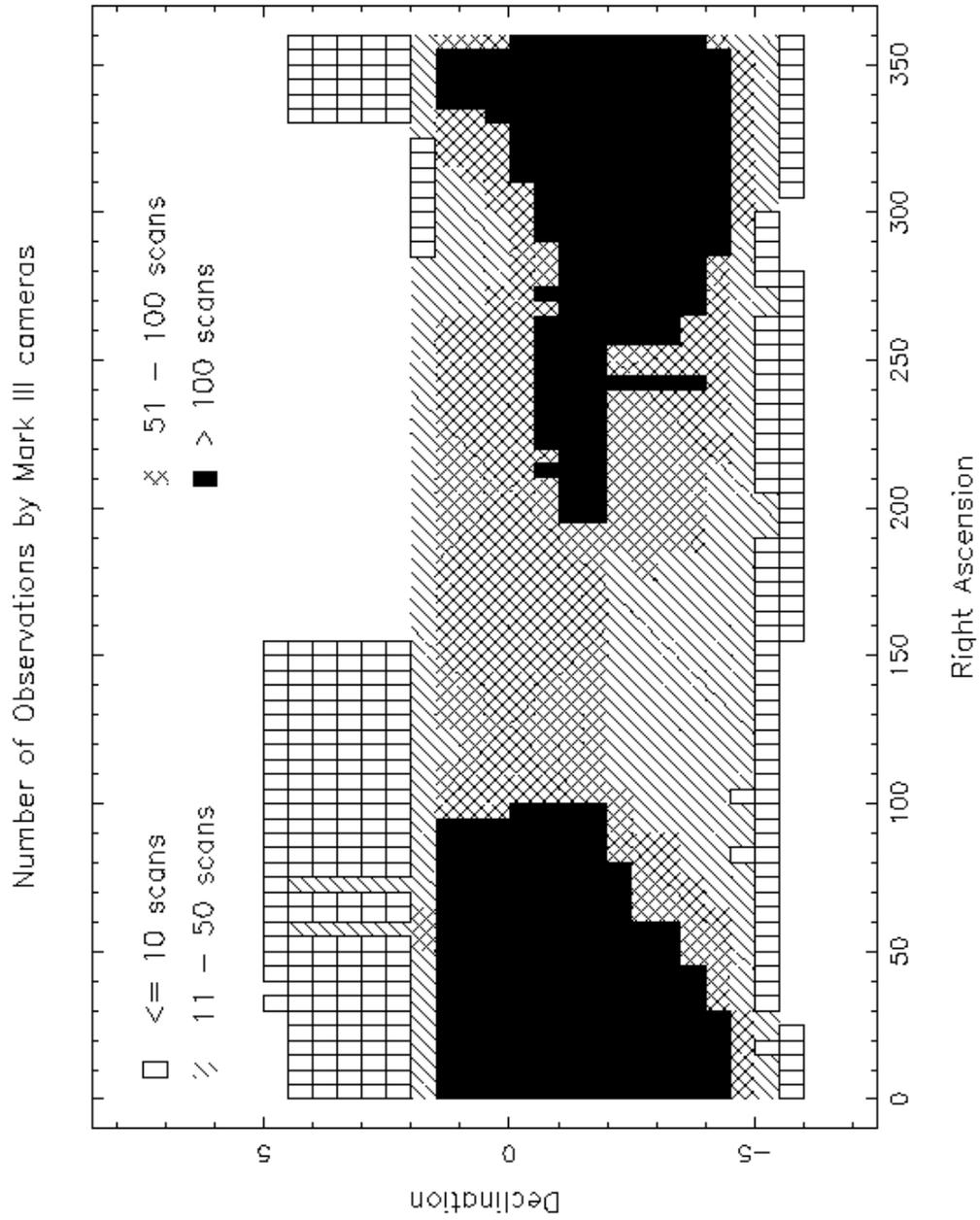}
\caption{
Number of observations by Mark III cameras as a function
of position on the sky.
}
\end{figure}

\begin{figure}
\figurenum{5}
\plotone{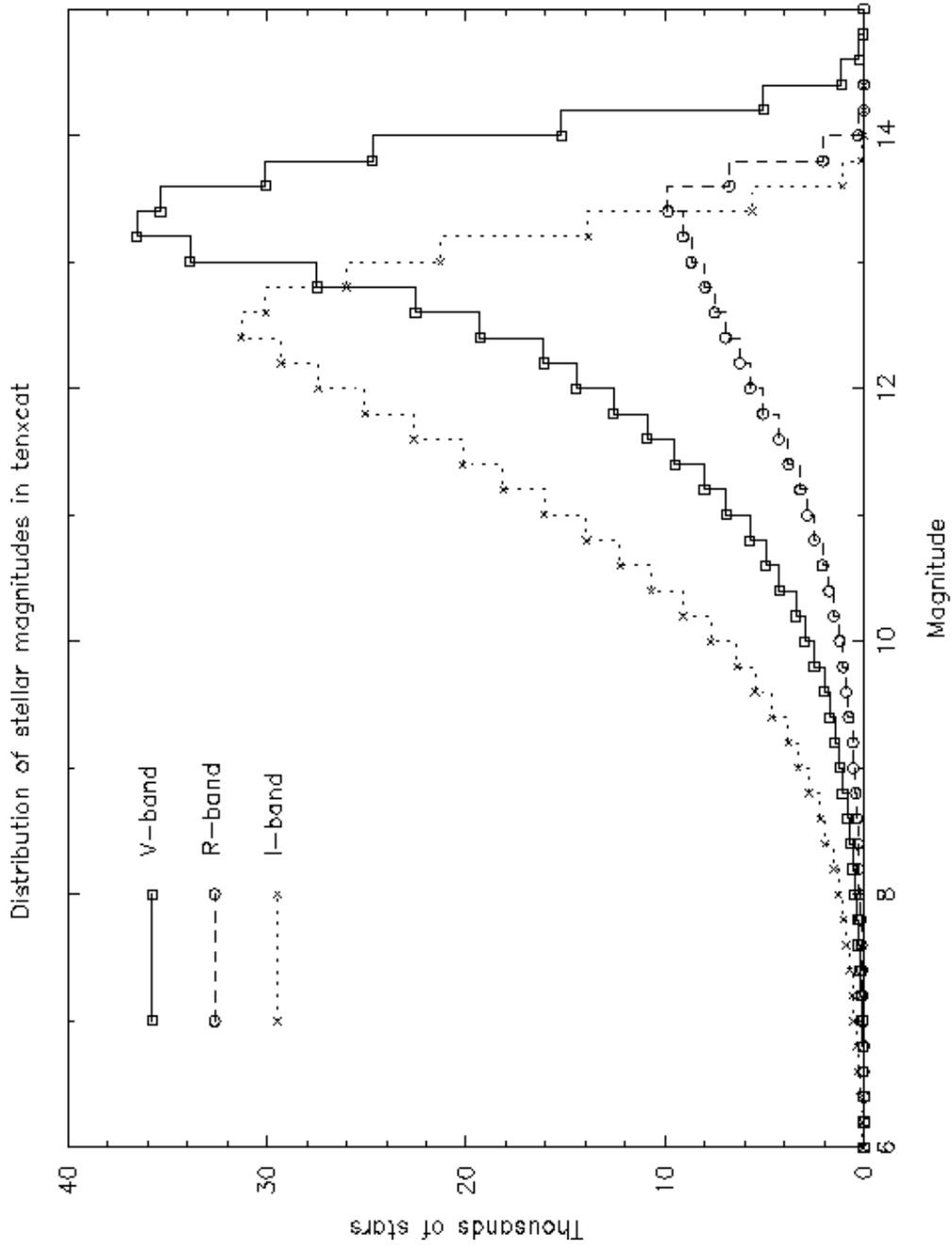}
\caption{
Distribution of magnitudes for stars in the {\it tenxcat} catalog.
}
\end{figure}

\begin{figure}
\figurenum{6}
\plotone{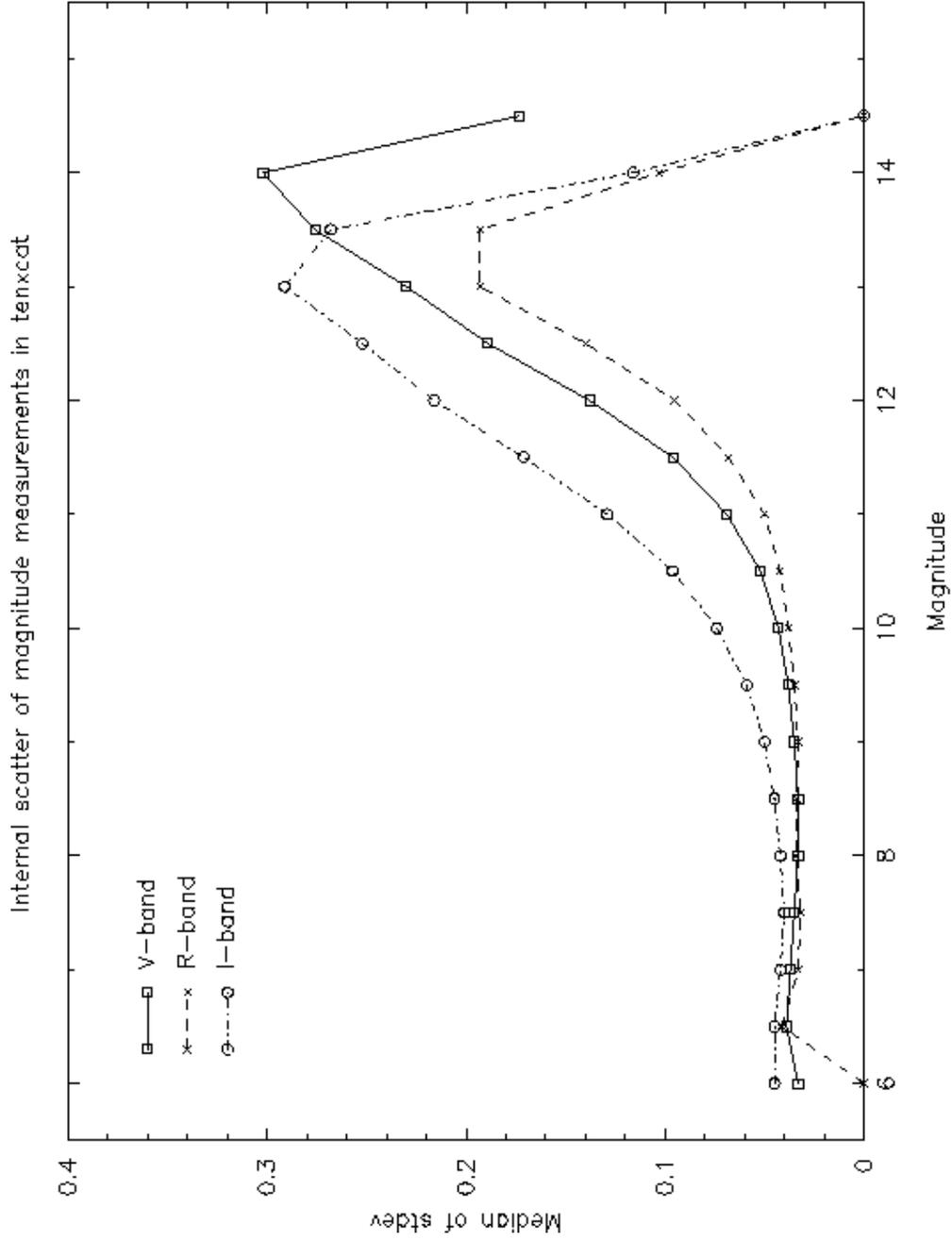}
\caption{
Median values of standard deviation from the mean for
magnitudes of stars in the {\it tenxcat} catalog.
}
\end{figure}

\end {document}